\documentclass[12pt]{article}
\usepackage{amssymb}
\usepackage{amsmath}
\usepackage{amsthm}
\usepackage{amsbsy}
\usepackage{amsfonts}
\usepackage[cp1251]{inputenc}
\usepackage[english,russian]{babel}
\usepackage[usenames]{color}
\newtheorem{thm}{Theorem}[section]

\newtheorem{prop}[thm]{Proposition}

\numberwithin{equation}{section}

\newcommand{\cM}{\mathcal {M}}
\newcommand{\cN}{\mathcal {N}}

\newcommand{\cA}{\mathcal {A}}

\newcommand{\cW}{\mathcal {W}}

\newcommand{\cE}{\mathcal {E}}

\newcommand{\cF}{\mathcal {F}}
\newcommand{\cG}{\mathcal {G}}
\newcommand{\cQ}{\mathcal {Q}}
\newcommand{\cD}{\mathcal {D}}

\newcommand{\cU}{\mathcal {U}}

\newcommand{\cV}{\mathcal {V}}

\newcommand{\fM}{\mathfrak {M}}

\newcommand{\fA}{\mathfrak {A}}

\newcommand{\fT}{\mathfrak {T}}

\newcommand{\bR}{\mathbb {R}}

\newcommand{\bC}{\mathbb {C}}

\begin{document}
\begin{center}
\begin{Large}
{\bf The Maxwell system in waveguides with several ends}
\end{Large}

\vspace{4ex}
\begin{large}
B. A. Plamenevskii, A. S. Poretckii
\end{large}
\footnote{This research is supported by the Chebyshev Laboratory  (Department of Mathematics and Mechanics, St. Petersburg State University)  under RF Government grant 11.G34.31.0026 and by  RFBR grant-12-01-00247a} 

\end{center}

A waveguide coincides with a domain $G$ in $\bR^3$ having finitely many cylindrical outlets to infinity; the boundary $\partial G$ is smooth. In $G$, we consider the stationary Maxwell system with spectral parameter $k \in \bR$ and identity matrices of dielectric and magnetic permittivity. The boundary $\partial G$ is supposed to be perfectly conductive. In the presence of charges and currents we investigate the solvability of the corresponding boundary value problem supplemented with "intrinsic"\, radiation conditions at infinity.  
For all $k$ in the continuous spectrum of the problem (including the thresholds and eigenvalues), we describe a basis in the space of continuous spectrum eigenfunctions, define the scattering matrix, and prove it is unitary. To this end, we extend the Maxwell system to an elliptic one and study the latter in detail. The information on the Maxwell boundary value problem  comes from that obtained for the elliptic problem. 
\section{Introduction. Formulation of the results}
Let $G$ be a domain in $\mathbb R^3$, coinciding outside a large ball with the union   $\Pi^1_+\cup \dots
\cup \Pi^N_+$ of finitely many non-overlapping semicylinders
$$
\Pi^q_+=\{(y^q, t^q): y^q\in \Omega^q, \, t^q>0\},
$$
where $(y^q, t^q)$ are local coordinates in $\Pi^q_+$ and  $\Omega^q$
is a bounded domain in $\mathbb R^2$. We consider the Maxwell system
\begin{eqnarray}\label{Max}\nonumber
i\,{\rm rot}\,u^2(x)- ku^1(x)&=& f^1(x), \\  -i\,{\rm
div}\,u^2(x)&=& h^1(x), \\ \nonumber -i\,{\rm rot}\,u^1(x)-
ku^2(x)&=& f^2(x), \\ \nonumber i\,{\rm div}\,u^1(x)&=& h^2(x)
\end{eqnarray}
in  $G$ with boundary conditions
\begin{eqnarray}\label{MaxBoun}
\nu (x) \times u^1(x)=0, \,\,\langle u^2(x), \nu (x)\rangle=0, \,\,
x\in
\partial G;
\end{eqnarray}
here $u^1$ and  $u^2$ are vector valued functions with three components (the electric and magnetic vectors respectively),  $k$ is a spectral parameter, $\nu$ is the outward normal to $\partial G$,
$\langle u, \nu \rangle$ are $\nu \times u$ are scalar and vector product, and  $f^j$, $h^j$ are given functions. The boundary conditions correspond to perfectly conductive boundary, which means that at $\partial
G$ there vanish the tangent component of $u^1$ and the normal one of $u^2$.  
The system (\ref{Max}) is overdetermined (eight equations and only six indeterminate functions). The compatibility conditions
\begin{eqnarray}\label{CC1} \nonumber
{\rm div} f^1 (x)-ikh^2(x)=0, \,\, x \in G, \\ 
{\rm div} f^2(x) + ikh^1(x)=0, \,\, x\in G,\\  \nonumber
\langle f^2 (x), \nu (x)\rangle =0, \,\, x\in \partial G  \nonumber
\end{eqnarray}
are necessary for the solvability of problem (\ref{Max}), (\ref{MaxBoun}).

Preparatory to describing the continuous spectrum eigenfunctions and the radiation conditions, we first
associate with problem (\ref{Max}), (\ref{MaxBoun}) some operator pencils and then define "$waves$"\, needed for asymptotic formulas. Let us write down (\ref{Max}), (\ref{MaxBoun}) in the form  
\begin{eqnarray}\label{MaxM}
\cM (D, k)\cU(x)= \cF (x), \, \quad x\in G; \quad u^1(x)_\tau=0, \,\quad u^2(x)_\nu=0, \,\quad x\in \partial G,
\end{eqnarray}
where $D = (D_1, D_2, D_3)$, $D_j= -i \partial /\partial x_j$,
$\mathcal U=(u^1, u^2)$, while  $u^1(x)_\tau$ and $u^2(x)_\nu$ denote the tangent 
and normal component of  $u^1(x)$ and $u^2(x)$.

Let us consider problem (\ref{MaxM}) in the cylinder $G:=\Omega \times \mathbb R
=\{x=(x_1, x_2, x_3): (x_1, x_2)\in \Omega, x_3\in \mathbb R\}$, where
$\Omega$ is a bounded domain in $\mathbb R^2$ with smooth boundary
$\partial \Omega$. For the vectors $\Phi = (\varphi, \psi)$ with components 
$\varphi$, $\psi$ in  $C^{\infty}(\bar \Omega; \mathbb C^3)$ satisfying the boundary conditions
$\varphi_\tau =0$, $\psi_\nu=0$ at $\partial \Omega$ (we assume here that the outward normal to $\partial \Omega$ is the vector $\nu =(\nu_1, \nu_2, 0)$),
we define an operator pencil  $\mathbb C \ni \lambda \mapsto \mathfrak{M} (\lambda, k)$
by the equality
\begin{equation}\label{pencilM}
\mathfrak M (\lambda, k)\Phi (x_1, x_2)=\exp{(-i\lambda x_3)}\mathcal M
(D, k)(\exp{(i\lambda x_3)}\Phi (x_1, x_2)).
\end{equation}
A number  $\lambda \in \bC$ is said to be an eigenvalue of the pencil $\fM(\cdot, k)$ if there exists a vector $\Phi \neq 0$ such that $\fM (\lambda, k)\Phi=0$; then $\Phi$ is called an eigenvector of the pencil. The eigenvalues of $\fM (\cdot, k)$ are resting on the coordinate axes of the complex plane, accumulate only at infinity, while the real axis contains at most finitely many eigenvalues. We assume that the domain $\Omega$ is 1-connected. A value $k \neq 0$ is called a threshold if $\lambda =0$ turns out to be an eigenvalue for $\fM (\cdot, k)$. The set of thresholds is infinite, symmetric about the coordinate origin, and accumulates only at infinity. The first threshold on the semiaxis $0<k<\infty$ is equal to $\sqrt{\mu_0}$, where $\mu_0$ is the minimal positive eigenvalue of the Neumann problem
\begin{eqnarray*}
\Delta u(x)+\mu u(x) =0, \quad x\in \Omega, \qquad  \partial_\nu u(x)=0, \quad x\in \partial \Omega,
\end{eqnarray*}
$\Delta$ being the Laplace operator in $\Omega$.  Let $k'$ and $k''$ are neighboring thresholds and let $0<k'<k''$. The sum of multiplicities $\kappa (k)$ of the real eigenvalues  $\lambda$ of the pencil $\fM(\cdot, k)$ is even and constant for  $|k|\in (k', k'')$; to every eigenvalue $\lambda$ there correspond only eigenvectors, while generalized eigenvectors do not exist.  We denote by $\lambda_0$ and $\Phi_0$ a real eigenvalue and a corresponding eigenvector of $\fM(\cdot, k)$. The function
\begin{equation}\label{homsol}
G\ni (y, t) \mapsto P(y, t)=\exp {(i\lambda_0 t)}\Phi_0 (y) 
\end{equation}
satisfies the homogeneous problem (\ref{MaxM}) in the cylinder $G=\{x=(y, t): y\in \Omega, t\in R \} $.
Each of the solutions is  a linear combination of TE- and TM-modes (see, e.g., \cite{J}). The dimension of the linear space of solutions of the form (\ref{homsol}) is equal to $\kappa (k)$. One can choose a basis in the space subject to certain orthogonality and normalization conditions so that  half of the basis consists of "waves outgoing to $+ \infty$"\,,  and the other half consists of "waves incoming from $+ \infty$"\,. 

Let us turn to the "general"\, domain $G$ with several cylindrical ends  $\Pi^1_+, \dots,
 \Pi^N_+$. We introduce the pencil $\fM^q (\cdot, k)$ in $\Pi^q=\Omega^q\times \bR$, $q=1, \dots, N$, 
denote by $\fT^q$ the set of thresholds for  $\fM^q $, and define 
$\fT:=\fT^1\cup \dots \cup \fT^N$. Assume now that $k', k''\in \fT$, $0<k'<k''$, while $(k', k'')\cap \fT=\emptyset$. We also suppose that $\varkappa (k):= \varkappa^1 (k)+\dots +\varkappa^N(k)$, where $k\in (k', k'')$ and $\varkappa^q(k)$ is the sum of multiplicities of the real eigenvalues of the pencil $\fM^q (\cdot, k)$. The value $\varkappa (k)$ is constant for $|k|\in (k', k'')$ and equal to an even number $2\Upsilon$. 

Denote by $\chi$ a function in $C^{\infty}(\bR)$ such that  $0\leq \chi (t)\leq 1$, $\chi (t)=1$ for $t>T$, and $\chi(t)=0$ for $t<T-1$, where  $T$ is sufficiently large number. In local coordinates of $\Pi^q_+$, introduce the function  $(y^q, t^q)\mapsto \chi (t^q)P(y^q, t^q)$,  $P$ being a solution of the form (\ref{homsol}) to the homogeneous problem (\ref{MaxM}) in $\Pi^q$. We can assume that ${\rm supp}\chi P\subset G\cap \Pi^q_+$ for a large $T$ in the definition of  $\chi$. Extending $\chi P$ by zero to $G$,
we obtain a smooth function given in $G$. Taking for $P$ in $\Pi^q$ a wave incoming from $+\infty$ (outgoing to $+\infty$), we obtain a function called incoming (outgoing) wave in the domain $G$. In such a way we define incoming waves $w^+_1, \dots, w^+_\Upsilon$ and outgoing waves 
$w^-_1, \dots, w^-_\Upsilon$ in $G$.

As before, assume that  $|k|\in (k', k'')$. A smooth bounded in  $G$  vector-valued function $\cU = (u^1, u^2)$ is called a continuous spectrum eigenfunction (CSE) of problem (\ref{Max}), (\ref{MaxBoun}) if  $\cU$ satisfies the homogeneous problem  (\ref{Max}), (\ref{MaxBoun}) and does not belong to the space $L_2 (G)$. 

Denote by $\rho_\delta$ a smooth positive function on $\overline G$, given on  $\Pi^q_+\cap G$ by
$\rho_\delta (y^q, t^q)=\exp {(\delta t^q)}$ with $q=1, \dots, N$ and a positive $\delta$. We choose  such a $\delta$ sufficiently small so that the strip $\{\lambda \in \bC: |\mathrm{Im}\lambda|\leq \delta \}$ contains no eigenvalues of the pencils $\fM^q(\cdot, k)$ except for the real ones. Introduce the space  $W^l_\delta (G)$ of functions in $G$ with norm 
$$
\|u; W^l_\delta (G)\|:=
(\sum_{|\alpha|=0}^l \int_G |D^\alpha (\rho_\delta u)|^2\,dx)^{1/2}
$$
for $l=0, 1, \dots $. A number $k$ is called an eigenvalue for problem (\ref{Max}), (\ref{MaxBoun}) if  there exists in $L_2 (G)$ a (smooth) solution to the homogeneous problem; as a rule, we do not distinct in notations the space of scalar and vector valued functions. Every eigenvalue is real; the dimension of eigenspace is always finite. The eigenvalues can accumulate only at infinity. Any eigenfunction turns out to be  in  $W^l_\delta (G)$ for each $l$. 

Denote by $\mathrm{ker}\cM (D, k)$ the eigenspace of problem  (\ref{Max}), (\ref{MaxBoun}) (possibly, trivial) corresponding to a number $k$. 
For the sake of formulation simplicity we suppose in Theorems \ref{t1.1} and \ref{t1.2}, as before, that  $k', k''\in \fT$, $0<k'<k''$, and  $(k', k'')\cap \fT=\emptyset$.
\begin{thm}\label{t1.1}
Let  $|k|\in (k', k'')$ and let  $2\Upsilon$ be the sum of multiplicities of all real eigenvalues of the pencils   $\fM^q(\cdot, k)$, $q=1, \dots, N$. Then there exist solutions $\cW_j^+ (\cdot, k)$, $j=1, \dots, \Upsilon$, to the homogeneous problem {\rm (\ref{Max}), (\ref{MaxBoun})} such that
$$
\cW_j^+ (\cdot, k) - w^+_j (\cdot, k) -\sum\limits_{q=1}^{\Upsilon}
s_{jq}(k)w_q^-(\cdot, k) \in W^l_\delta (G),$$
where $l=1,2, \dots$
If  $k$ is not an eigenvalue of problem {\rm (\ref{Max}), (\ref{MaxBoun})}, then the solution $\cW_j^+ (\cdot, k)$ with above property is unique and   $\cW_1^+(\cdot, k), \dots, \cW_\Upsilon^+ (\cdot, k)$ form a basis in the space of CSE corresponding to $k$. If $k$ turns out to be an eigenvalue of problem {\rm (\ref{Max}), (\ref{MaxBoun})},  then $\cW_j^+ (\cdot, k)$ is determined up to an arbitrary term in $\mathrm{ker}\cM (D, k)$ and any 
$\cW_1^+(\cdot, k), \dots, \cW_\Upsilon^+ (\cdot, k)$
form a basis modulo $\mathrm{ker}\cM (D, k)$
in the space of CSE.

The matrix $$s (k)=(s_{jq}(k))_{j,\, q=1}^\Upsilon$$ is unitary. It is independent of the choice of $\cW_j^+(\cdot, k)$ in the case that 
$k$ is an eigenvalue of problem {\rm (\ref{Max}), (\ref{MaxBoun})}.
\end{thm}
The theorem can be generalized for the thresholds as well. The matrix  $s$ is called the scattering matrix. It is defined for $k$ satisfying $k^2\geq \mu_{\fM}$, where
$\mu_{\fM}= \min \{\mu^1_0, \dots, \mu^N_0\} $ and $(\mu^j_0)^{1/2}$ is the first positive threshold for the pencil
$\fM^j$ in $\Omega^j$. In fact, there is no wave for $k^2< \mu_{\fM}$ that could transfer energy.  

The incoming and outgoing waves can "exchange roles"\,. In particular, there exist solutions $\cW^-_j(\cdot, k)$ to the homogeneous problem (\ref{Max}), (\ref{MaxBoun}) such that 
$$
\cW_j^- (\cdot, k) - w^-_j (\cdot, k) -\sum\limits_{q=1}^{\Upsilon}
t_{jq}(k)w_q^+(\cdot, k) \in W^l_\delta (G).
$$
The matrices $(s_{jq}(k))_{j,\, q=1}^\Upsilon$ and  $(t_{jq}(k))_{j,\, q=1}^\Upsilon$ are mutually inverse.

The solvability of the problem  (\ref{Max}), (\ref{MaxBoun}) supplemented with intrinsic radiation conditions is established by the following theorem.
\begin{thm}\label{t1.2}
Let $|k|\in (k', k'')$ and let $\zeta_1, \dots, \zeta_m$ be a basis in the space ${\rm ker} \cM (D, k)$ of eigenvectors of problem {\rm (\ref{Max}), (\ref{MaxBoun})}. We also assume that 
$\cF =(f^1, h^1, f^2, h^2)$ is in $W^{l-1}_\delta (G, \bC^8)$, ($l\geq 1$), satisfies the compatidility
conditions {\rm (\ref{CC1})} and moreover 
 $(f, \zeta_j)_G=0$ for $j=1, \dots, m$, where $f=(f^1, f^2)$  and $(\cdot, \cdot)_G$ is the inner product on $L_2 (G)$. Then there exists a solution  $\cU =(u^1, u^2)$ with the radition conditions
\begin{equation*}
 \cV:=\cU-c_1 w^-_1 - \dots  c_\Upsilon w^-_\Upsilon \in W^l_\delta (G; \bC^6),
\end{equation*}
where $c_j=i(f, \cW^-_j)_G$. Such a solution $\cU$ is determined up to an arbitrary term in 
 ${\rm ker}\, \cM (D, k)$ and there holds the inequality 
 \begin{equation}\label{ElR3}
\|\cV; W^l_\delta (G; \bC^6)\| +|c_1|+ \dots + |c_\Upsilon| \leq {\rm const} (\|\cF; W^{l-1}_\delta (G,  \bC^8)\|
+\|\rho_\delta \cV; L_2 (G, \bC^6)\|).
 \end{equation} 
A solution $\cU_{\,0}$ that satisfies the additional conditions $(\cU_{\,0}, \zeta_j)_G=0$ 
is unique and there holds the estimate
{\rm (\ref{ElR3})} with right-hand changed for ${\rm
const}\|\cF; W^{l-1}_\delta (G, \bC^8 ) \|$.
\end{thm}
The study of problem  (\ref{Max}), (\ref{MaxBoun}) begins with extension of the overdetermined Maxwell system to an elliptic system. To this end we use the orthogonal extension method suggested by I.S. Gudovich,
S.G. Krein, and I.M. Kulikov (see e.g. \cite{DS}, \cite{GK} and references there). As a result, there arises an elliptic boundary value problem self-adjoint with respect to a Green formula.  The general problems of this type in domains with cylindrical outlets to infinity were studied in \cite{NP}. In particular, the intrinsic radiation conditions were described, the solvability of the boundary value problem with those radiation conditions was established,  the unitary scattering matrix was introduced. When analyzing the obtained elliptic problem, we clarify its specific properties coming from the Maxwell system. To this end we investigate in detail the operator pencil generated by the elliptic problem. Then we derive the information on the Maxwell system from that obtained for the elliptic one. 

From numerous mathematical works devoted to the Maxwell system in waveguides we set off two lines of investigation. One of the lines is related to the Wiener-Hopf technique and the mode matching method. Surveys of the methods are given in  \cite{W2}, \cite{ML}. The other line is presented in \cite{BDS1} (see also references therein). The methodology in these works is connected with 
cylindrical waveguides and dielectric and magnetic permittivity independent of the axial variable.

We use neither the methods nor the results of the works mentioned in the preceding paragraph. The  elliptic extension of Maxwell system provides all the advantages of elliptic situation, in particular, the possibility of localization, freedom in choosing waveguide geometry. In this paper, we consider waveguides with identity matrices of dielectric and magnetic permittivity; in another paper, we are going to show that our approach also suggests reasonably wide freedom in choosing  waveguide medium.
\section{Augmented Maxwell system}
\subsection{Elliptic boundary value problem}
We now pass on to the "orthogonal extension"\, of system (\ref{Max}) (\cite{DS}, see also \cite{GK}). Namely, in the domain $G$ we introduce the boundary value problem 
\begin{eqnarray}\label{ExtMax}\nonumber
i\,{\rm rot}\,u^2(x)+i\,\nabla a^2(x)- ku^1(x)&=& f^1(x),\\
 -i\,{\rm div}\,u^2(x)-k a^1(x)&=& h^1(x), \\
\nonumber
-i\,{\rm rot}\,u^1(x)- i\,\nabla a^1(x) - ku^2(x)&=& f^2(x),\\
i\,{\rm div}\,u^1(x)- ka^2(x)&=& h^2(x)\nonumber
\end{eqnarray}
with boundary conditions
\begin{eqnarray}\label{ExtBound}
\nu (x) \times u^1(x)=g^1(x), \,\,\langle u^2(x), \nu (x)\rangle
=g^2(x), \,\,a^2(x)=g^3(x), \,\, x\in \partial G;
\end{eqnarray}
here $u^1, u^2$ are vector valued functions with three components and  $a^1, a^2$ stand for scalar functions in $G$.
Problem {\rm (\ref{ExtMax}), (\ref{ExtBound})} is elliptic. Rewrite it in the form
\begin{eqnarray}\label{Probl}
\mathcal A (D, k)\mathcal U (x)& =& \mathcal F (x), \,\, x\in G, \\
\nonumber \mathcal B\mathcal U (x)& = &\mathcal G (x), \,\, x\in
\partial G,
\end{eqnarray}
where $D = (D_1, D_2, D_3)$, $D_j= -i \partial /\partial x_j$,
$\mathcal U=(u^1, a^1, u^2, a^2)$. The Green formula holds
\begin{equation}\label{Green}
(\mathcal A (D, k)\mathcal U, \mathcal V)_G+(\mathcal B \mathcal U,
\mathcal Q \mathcal V)_{\partial G} = (\mathcal U, \mathcal A (D,
k)\mathcal V)_G + (\mathcal Q \mathcal U, \mathcal B \mathcal
V)_{\partial G},
\end{equation}
with $\mathcal U=(u^1, a^1, u^2, a^2)$, $\mathcal V =(v^1, b^1, v^2,
b^2)$, and
\begin{eqnarray*}
\mathcal B \mathcal U = (\nu \times u^1, \langle u^2, \nu \rangle,
a^2), \,\,\, \mathcal Q \mathcal V = (-iv^2, -ib^1, \langle iv^1,
\nu \rangle).
\end{eqnarray*}
The operator of problem {\rm (\ref{Probl})} is self adjoint with respect to the Green formula
{\rm (\ref{Green})}.
\subsection{Elliptic and Maxwell operator pencils}\label{s3.1}
We consider the operator  $\{\mathcal A (D), \mathcal
B\}$ of problem  (\ref{Probl}) in the cylinder  $\Omega \times \mathbb R
=\{x=(x_1, x_2, x_3): (x_1, x_2)\in \Omega, x_3\in \mathbb R\}$, where 
$\Omega$ is a bounded domain in $\mathbb R^2$ with smooth boundary
$\partial \Omega$. Let  $\Phi =(\varphi, \alpha, \psi,
\beta)$ be a vector with components $\varphi$, $\psi$ in $C^{\infty}(\bar \Omega
; \mathbb C^3)$ and $\alpha$, $\beta$ in $C^{\infty}(\bar \Omega;
\mathbb C)$ satisfying
\begin{equation}\label{homboundcond}
\mathcal B \Phi =(\nu \times \varphi, \langle \psi, \nu \rangle,
\beta)=0
\end{equation}
on $\partial \Omega$, where $\nu$ is the outward normal to $\partial
\Omega$; we assume that  $\nu$ is of the form
$(\nu_1, \nu_2, 0)$. Denote by $u_\tau$ and $u_\nu$ the tangent and normal components of
$u$ on $\partial \Omega$ and rewrite
(\ref{homboundcond}) as
$$
\varphi_\tau=0, \,\, \psi_\nu =0, \,\, \beta|\partial\Omega =0.
$$
For the vectors $\Phi$ with such properties we define the operator pencil $\mathbb C \ni \lambda \mapsto \mathfrak A (\lambda)$,

\begin{equation}\label{pencil}
\mathfrak A (\lambda)\Phi (x_1, x_2)=\exp{(-i\lambda x_3)}\mathcal A
(D)(\exp{(i\lambda x_3)}\Phi (x_1, x_2)).
\end{equation}
For the usual operations $\nabla$, ${\rm rot}$, ${\rm div}$, and $\Delta$ in
$\Omega \times \mathbb R$, introduce in  $\Omega$ the operations
$\nabla (\lambda)$, ${\rm rot}(\lambda)$, ${\rm div} (\lambda)$ and
$\Delta (\lambda)$ by
\begin{eqnarray*}\nabla (\lambda)
\alpha (x_1, x_2) = \exp{(-i\lambda x_3)} \nabla \,(\exp{(i\lambda
x_3)} \alpha (x_1, x_2)),\\
{\rm rot}(\lambda)\varphi (x_1, x_2) = \exp{(-i\lambda x_3)}{\rm
rot} \, (\exp{(i\lambda x_3)}\varphi (x_1, x_2)),
\end{eqnarray*}
etc. Formulas for the usual operations can  immediately be modified for the operations with parameter.  For example, from
${\rm rot}\,{\rm rot}=\nabla\,{\rm div} -\Delta$ it follows that
${\rm rot}(\lambda)\,{\rm rot}(\lambda)=\nabla (\lambda)\,{\rm
div}(\lambda) -\Delta (\lambda)$. For $\varphi$, $\psi$ in
$C^{\infty}(\bar \Omega ; \mathbb C^3)$ and $\alpha$ in
$C^{\infty}(\bar \Omega ; \mathbb C)$ we have
\begin{eqnarray}\label{ort1}
(\nabla (\lambda)\alpha, \varphi)_\Omega &=&(\alpha, \langle
\varphi, \nu \rangle)_{\partial \Omega} -(\alpha, {\rm div} (\bar \lambda)\varphi)_\Omega, \\
\label{ort2} ({\rm rot}(\lambda)\varphi, \psi)_\Omega &=&(\varphi,
\psi \times \nu)_{\partial \Omega} +(\varphi, {\rm rot}(\bar
\lambda)\psi)_\Omega.
\end{eqnarray}
Denote by $H^l(\Omega; \mathbb C^8)$, $l= 0, 1, \dots$,
the space of vectors with eight components in the Sobolev space 
$H^l(\Omega; \mathbb C)$ of functions in $\Omega$.
We write the elements $\Phi \in H^l (\Omega; \mathbb C^8)$ as 
$\Phi=(\varphi, \alpha, \psi, \beta)$, where $\varphi, \psi \in H^l
(\Omega; \mathbb C^3)$ and $\alpha, \beta \in H^l(\Omega; \mathbb C)$.
For $l=1, 2, \dots$  set
\begin{equation}\label{DH}
\cD H^l(\Omega)=\{\Phi \in H^l (\Omega; \mathbb C^8):
\varphi_\tau=0, \psi_\nu =0, \beta|\partial \Omega =0\}.
\end{equation}
Let us consider the operator  $\fA (\lambda)$ given by (\ref{pencil}) on
the domain $\cD H^l(\Omega)$. According to the general theory of elliptic operator pencils (see \cite{AV}), for all 
$\lambda \in \mathbb C$ with the exception of some isolated points, the mapping  $\fA (\lambda): \cD H^l(\Omega)\to
H^{l-1}(\Omega; \mathbb C^8)$is an isomorphism. The mentioned isolated points are the eigenvalues of the pencil $\lambda
\mapsto \fA (\lambda)$ of finite algebraic multiplicity. The components of eigenvectors and 
generalized eigenvectors are smooth in 
$\overline \Omega$. For $\cU =(\varphi, \alpha, \psi, \beta) \in \cD
H^l(\Omega)$ from (\ref{ExtMax}) and  (\ref{pencil}) it follows that
\begin{equation}\label{pencilExplit}
\fA (\lambda): \left( \begin{array}{c}\varphi \\ \alpha \\ \psi \\
 \beta \end{array} \right) \mapsto \left(
\begin{array}{c} i\,{\rm rot (\lambda)}\,\psi +i\,\nabla (\lambda) \beta -
k \varphi  \\  -i\,{\rm div} (\lambda) \,\psi -k \alpha \\
-i\,{\rm rot} (\lambda)\,\varphi - i\,\nabla (\lambda) \alpha - k
\psi
\\ i\,{\rm div} (\lambda)\,\varphi- k \beta
\end{array} \right).
\end{equation}
The pencil  $\fA$ is called elliptic and its restriction to 
$\{\cU \in \cD H^l(\Omega): \cU =(\varphi, 0, \psi, 0)\}$ will be called the Maxwell pencil and denoted by $\fM$.  The number  $\lambda_0$ is an eigenvalue of the pencil $\fM (\cdot, k)$ if there exists a smooth nonzero vector $\Phi =(\varphi, 0, \psi, 0)$ that is subject to the boundary conditions
$\varphi_\tau=0, \psi_\nu =0$ on $\partial \Omega$ and satisfies 
$\fM (\lambda_0, k)\Phi=0$.
\subsection{Eigenvalues and eigenvectors of the pencils $\fA$ and $\fM$}
\begin{prop}\label{tspect1}
Let  $\lambda$ be an eigenvalue of the pencil $\fA (\cdot, k)$ and 
$(\varphi, \alpha, \psi, \beta)$ an eigenvector corresponding to the eigenvalue, 
$\varphi =(\varphi_1, \varphi_2, \varphi_3)$ and 
$\psi= (\psi_1, \psi_2, \psi_3)$. Assume that
$k^2-\lambda^2\neq 0$. Then
\begin{eqnarray}\label{HAlp}
\Delta (\lambda)\alpha +k^2 \alpha&=&0\,\, {\text in} \,\,
\Omega,\qquad \partial_\nu \alpha =0 \,\,\,{\text
{on}}\,\,\,\partial \Omega, \\
\label{HBet} \Delta (\lambda) \beta +k^2 \beta&=&0\,\, {\text in}
\,\, \Omega,
\qquad \beta =0 \,\,\,{\text {on}}\,\,\,\partial \Omega, \\
\label{HPh} \Delta (\lambda)\varphi_3+k^2 \varphi_3&=&0\,\, {\text
in} \,\, \Omega,\qquad  \varphi_3 =0 \,\,\,{\text
{on}}\,\,\,\partial \Omega, \\
\label{HPs} \Delta (\lambda)\psi_3+k^2 \psi_3&=&0\,\, {\text in} \,\,
\Omega, \qquad \partial_\nu \psi_3 =0 \,\,\,{\text
{on}}\,\,\,\partial \Omega,
\end{eqnarray}
while  $\varphi_j$, $\psi_j$ for  $j=1, 2$ are defined by
\begin{eqnarray}\label{expr}
\varphi_1& =& (k^2-\lambda^2)^{-1}[i \lambda \partial_1 \varphi_3
+i k\partial_2 \psi_3 -i\lambda \partial_2 \alpha +i k\partial_1
\beta],\nonumber \\
\varphi_2&=&(k^2-\lambda^2)^{-1}[i \lambda
\partial_2 \varphi_3 - i k \partial_1 \psi_3  +i \lambda \partial_1 \alpha
+i k \partial_2 \beta ], \\
\psi_1&=&(k^2-\lambda^2)^{-1}[-ik\partial_2 \varphi_3 +i\lambda
\partial_1 \psi_3 - i k \partial_1 \alpha - i \lambda \partial_2
\beta], \nonumber \\
\psi_2&=&(k^2-\lambda^2)^{-1}[i k \partial_1 \varphi_3 + i \lambda
\partial_2 \psi_3 - i k \partial_2 \alpha +i \lambda \partial_1
\beta]. \nonumber
\end{eqnarray}
Conversely, any nonzero vector $(\varphi,
\alpha, \psi, \beta)$ with components satisfying
{\rm (\ref{HAlp}) -- (\ref{HPs})} and {\rm (\ref{expr})} is an eigenvector of
$\fA (\cdot, k)$ corresponding to $\lambda$.

If a number $\lambda$ (such that $k^2-\lambda^2\neq 0$) is an eigenvalue for one of the  
pencils $\fA (\cdot, k)$ and $\fM (\cdot,
k)$, then it is an eigenvalue for the other one; moreover, $\lambda$ turns out to be an eigenvalue of the pencils if and only if it is an eigenvalue for at least one of problems
{\rm (\ref{HAlp}) and (\ref{HBet})}. There hold the equalities 
$$
\varkappa_{\fA}(\lambda, k)=2\varkappa_{\fM}(\lambda,
k)=2\varkappa_{\cD}(\lambda, k)+2\varkappa_{\cN}(\lambda, k),
$$
where  $\varkappa_{\fA}(\lambda, k)$ and $\varkappa_{\fM}(\lambda, k)$
are the geometric multiplicities of the eigenvalue  $\lambda$ for the pencils
$\fA (\cdot, k)$ and $\fM (\cdot, k)$, while  $\varkappa_{\cN}(\lambda, k)$
and $\varkappa_{\cD}(\lambda, k)$ are those for problems  {\rm (\ref{HAlp})} and
{\rm (\ref{HBet})}.
\end{prop} 
The case $k^2-\lambda^2=0$ has been described in the next proposition.
\begin{prop}\label{tspect2}
Let $\Omega$ be a 1-connected domain. Then:

1. If $\lambda^2 = k^2 \neq 0$, then $\lambda$ is an eigenvalue of  $\fA (\cdot, k)$ and the corresponding  
eigenspace is one-dimensional and spanned by the vector   $\Phi =(\varphi, \alpha, \psi,
\beta)$ with components
$$
\varphi=0, \qquad \alpha = {\rm const}\neq 0, \qquad
\psi_1=\psi_2=0, \qquad \psi_3=(\lambda/k)\alpha, \qquad \beta=0.
$$
The vector $\Phi$ does not belong to the domain  $\fM (\cdot,
k)$, and  $\lambda$ is not an eigenvalue for  $\fM (\cdot, k)$.

2. If  $k=0$, then  $\lambda =0$ is an eigenvalue of $\fA
(\cdot, k)$ with eigenspace spanned by the vectors $\widehat \Phi = (\hat \varphi, \hat \alpha,
\hat \psi, \hat \beta)$and $\widetilde \Phi =(\tilde \varphi, \tilde
\alpha, \tilde \psi, \tilde \beta)$, where
$$\hat
\varphi=0, \qquad \hat \alpha ={\rm const}\neq 0, \qquad \hat \psi
=0, \qquad \hat \beta =0,
$$
$$
 \tilde \varphi =0, \qquad \tilde \alpha=0, \qquad
\tilde \psi_1=\tilde \psi_2 =0, \qquad \tilde \psi_3={\rm const}\neq
0, \qquad \tilde \beta =0.
$$
The vector $\widehat \Phi$ does not belong to the domain of  $\fM
(\cdot, k)$, while  $\widetilde \Phi$  is an eigenvector for
$\fM (\cdot, k)$. 
\end{prop}
The generalized eigenvectors of $\fA (\cdot, k)$ exist if and only if $k$ is a threshold, i.e. $\lambda=0$ is an eigenvalue  of $\fA (\cdot, k)$ while $k\neq 0$. We do not dwell on describing such vectors.
\subsection{Continuous spectrum eigenfunctions. Scattering matrix} 
Assume that $k$ is not a threshold for the elliptic pencils $\fA^1, \dots \fA^N$. For every $q=1, \dots, N$ in the cylinder $\Pi^q =\Omega^q\times \mathbb R$, we define solutions of the form (\ref{homsol}) to the homogeneous elliptic problem (\ref{ExtMax}), (\ref{ExtBound}) (to this end we can use Propositions \ref{tspect1} and \ref{tspect2}). The dimension of linear space of such solutions in $\Pi^q$ is equal to
the sum $\varSigma_{\fA}^q (k)$ of all real eigenvalue multiplicities for $\fA^q (\cdot, k)$; the number $\varSigma_{\fA}^q (k)$ is even. In the space there exists a basis satisfying proper orthogonality and
normalization conditions so that one half of the basis consists of of "waves outgoing to $+ \infty$"\,,  and the other half consists of "waves incoming from $+ \infty$"\,. The waves in the domain $G$
\begin{eqnarray}
v^+_j, \quad  v^-_j, \qquad j=1, \dots, T:= (\varSigma^1_{\fA}(k)+ \dots + \varSigma^N_{\fA}(k))/2, 
\end{eqnarray}
 can now be introduced like those in Section 1 after (\ref{homsol}). 

 A smooth bounded in  $G$  vector-valued function $\cU = (u^1, a^1, u^2, a^2)$ is called a continuous spectrum eigenfunction (CSE) of problem (\ref{ExtMax}), (\ref{ExtBound}) if  $\cU$ satisfies the homogeneous problem  (\ref{ExtMax}), (\ref{ExtBound}) and does not belong to the space $L_2 (G)$.  A number $k$ is called an eigenvalue for problem (\ref{ExtMax}), (\ref{ExtBound}) if  there exists in $L_2 (G)$ a (smooth) solution to the homogeneous problem. Every eigenvalue is real; the dimension of eigenspace is always finite. The eigenvalues can accumulate only at infinity. Any eigenfunction turns out to be  in  $W^l_\delta (G)$ for each $l$. Denote by $\mathrm{ker}\cA (D, k)$ the eigenspace of problem  (\ref{ExtMax}), (\ref{ExtBound}) (possibly, trivial) corresponding to a number $k$. 
\begin{thm} \label{t2.3}
Let  $2T$ be the sum of all real eigenvalue multiplicities  of the pencils   $\fA^q(\cdot, k)$, $q=1, \dots, N$. Then there exist solutions $\cV_j^+ (\cdot, k)$, $j=1, \dots, T$, to the homogeneous problem {\rm (\ref{ExtMax}), (\ref{ExtBound})} such that
$$
\cV_j^+ (\cdot, k) - v^+_j (\cdot, k) -\sum\limits_{q=1}^{T}
\sigma_{jq}(k)w_q^-(\cdot, k) \in W^l_\delta (G),$$
where $l=1,2, \dots$
If  $k$ is not an eigenvalue of problem {\rm (\ref{ExtMax}), (\ref{ExtBound})}, then the solution $\cV_j^+ (\cdot, k)$ with above property is unique and   $\cV_1^+(\cdot, k), \dots, \cV_T^+ (\cdot, k)$ form a basis in the space of CSE corresponding to $k$. If $k$ turns out to be an eigenvalue of problem {\rm (\ref{ExtMax}), (\ref{ExtBound})},  then $\cV_j^+ (\cdot, k)$ is determined up to an arbitrary term in $\mathrm{ker}\cA (D, k)$ and any 
$\cV_1^+(\cdot, k), \dots, \cV_T^+ (\cdot, k)$
form a basis modulo $\mathrm{ker}\cA (D, k)$
in the space of CSE.

The matrix $$\sigma (k)=(\sigma_{jq}(k))_{j,\, q=1}^T$$ is unitary. It is independent of the choice of $\cV_j^+(\cdot, k)$ in the case that 
$k$ is an eigenvalue of problem {\rm (\ref{ExtMax}), (\ref{ExtBound})}.
There exist solutions $\cV^-_j(\cdot, k)$ to the homogeneous problem (\ref{ExtMax}), (\ref{ExtBound}) such that 
\begin{eqnarray}\label{tau}
\cV_j^- (\cdot, k) - v^-_j (\cdot, k) -\sum\limits_{q=1}^{T}
\tau_{jq}(k)v_q^+(\cdot, k) \in W^l_\delta (G).
\end{eqnarray}
The matrices $(\sigma_{jq}(k))_{j,\, q=1}^T$ and  $(\tau_{jq}(k))_{j,\, q=1}^T$ are mutually inverse.
\end{thm}
This theorem can be generalized for any real $k$ (including the thresholds), the  $\sigma (k)$ and $\tau (k)$ are defined  for all $k\in \bR$, i.e., the  continuous spectrum of problem  (\ref{ExtMax}), (\ref{ExtBound}) coincides with $\bR$.
\subsection{Radiation principle}\label{ss5.3} 
\begin{thm}\label{Radiation}
Let $z_1, \dots, z_d$ be a basis in ${\rm
ker}\,\cA (D, k)$ and  $\{\cF, \cG\}\in W^{l-1}_\delta (G)\times
W^{l-1/2}_\delta (\partial G)$, while
$$(\cF, z_j)_G +(\cG, \cQ z_j)_{\partial G}=0, \quad j=1, \dots, d,$$
where $\cQ$ is the operator in the Green formula {\rm (\ref{Green})}. Then
there exists a solution $\cU$ of the problem {\rm (\ref{ExtMax}), (\ref{ExtBound})}
determined up to an arbitrary term in ${\rm ker}\,\cA (D, k)$ such that 
\begin{equation}\label{Radia1}
\cV:= \cU -c_1v_1^- - \dots -c_T v_T^-  \in W^l_\delta (G),
\end{equation}
where 
\begin{equation}\label{RadCoef}
c_j= i(\cF, \cV_j^-)_G + i(\cG, \cQ \cV_j^-)_{\partial G}, \quad
j=1, \dots, T.
\end{equation}
There holds the inequality
\begin{eqnarray}\label{Radia2}
\|\cV; W^l_\delta (G)\|+|c_1|+\dots + |c_ T|\\ \nonumber \leq
{\rm const}(\|\cF; W^{l-1}_\delta (G) \|+\|\cG; W^{l-1/2}_\delta
{\partial G}\|+ \|\rho_\delta \cV; L_2 (G)\|).
\end{eqnarray}
A solution  $\cU_{\,0}$ subject to  $(\cU_{\,0},
z_j)_G=0$, $j=1, \dots, d$, is unique and  satisfies 
{\rm (\ref{Radia2})} with right-hand side replaced by  ${\rm
const}(\|\cF; W^{l-1}_\delta (G) \|+\|\cG; W^{l-1/2}_\delta
{\partial G}\|)$.
\end{thm}
\section{Coming back to the non-augmented Maxwell system}
To simplify formulations, we suppose that $k\neq 0$ and $k$ is not a threshold.
We will consider the elliptic system
\begin{eqnarray}\label{ExtMaxA}\nonumber
i\,{\rm rot}\,u^2(x)+i\,\nabla a^2(x)- ku^1(x)&=& f^1(x),\\
 -i\,{\rm div}\,u^2(x)-k a^1(x)&=& h^1(x), \\
\nonumber
-i\,{\rm rot}\,u^1(x)- i\,\nabla a^1(x) - ku^2(x)&=& f^2(x),\\
i\,{\rm div}\,u^1(x)- ka^2(x)&=& h^2(x)\nonumber
\end{eqnarray}
in $G$ with homogeneous boundary conditions 
\begin{eqnarray}\label{ExtBoundA}
\nu (x) \times u^1(x)=0, \,\,\langle u^2(x), \nu (x)\rangle =0,
\,\,a^2(x)=0, \,\, x\in \partial G.
\end{eqnarray}
If $\cF =(f^1, h^1, f^2, h^1)$ belongs to $W^{l-1}_\delta(G)$ with $l\geq 2$, then the compatibility conditions 
(\ref{CC1}) can be understood directly. In fact, such conditions can be interpreted in some generalized form for $l=1$ as well.  

\begin{prop}\label{t3.1}
Assume that the vector   $\cF =(f^1, h^1, f^2, h^2)$ belongs to 
$W^{l-1}_\delta (G)$ and is subject to compatibility conditions {\rm (\ref{CC1})}. Let $\cU =(u^1, a^1, u^2, a^2)$ satisfy the problem {\rm (\ref{ExtMaxA})}, {\rm (\ref{ExtBoundA})} and the radiation condition {\rm (\ref{Radia1})}. 
Then $a^1$ is a solution to the problem
\begin{equation}\label{PNG}
(\Delta +k^2) a^1(x)=0, \quad x \in G, \quad \partial_\nu a^1(x)=0,
\quad x\in \partial G,
\end{equation}
and satisfies the intrinsic radiation conditions defined for the problem {\rm (\ref{PNG})}, while $a^2$ is a solution to the problem
\begin{equation}\label{PDG}
(\Delta +k^2) a^2(x)=0, \quad x \in G, \quad  a^2(x)=0, \quad x\in
\partial G,
\end{equation}
and satisfies the intrinsic radiation conditions defined for  problem {\rm (\ref{PDG})}. Thus $a^1$ ($a^2$)
can be nonzero only if it is an eigenfunction of problem {\rm(\ref{PNG})} (problem {\rm (\ref{PDG})}).
\end{prop}
Let us here restrict ourselves to considering the case that $k$ is an eigenvalue of neither of problems (\ref{PNG}) and (\ref{PDG}). Then the solution $\cU$ in Proposition \ref{t3.1} in fact satisfies (\ref{ExtMaxA}) and  (\ref{ExtBoundA}) with $a^1 = a^2 =0$. It remains to discuss the radiation conditions
(\ref{Radia1}).

The collection of waves in G needed for problem (\ref{ExtMaxA}), (\ref{ExtBoundA}) in Theorems \ref{t2.3} and \ref{Radiation}   consists of the two parts 
\begin{equation}\label{WaveSet1}
E=\{e^{\pm}_j\}_{j=1}^\Upsilon \quad {\text{and}} \quad
\Gamma=\{\gamma^{\pm}_j\}_{j=1}^{\Upsilon+N}.
\end{equation}
The outgoing and incoming waves  $e^{\pm}_j$ are generated by eigenvectors of $\fM^q (\cdot, k)$, while the waves $\gamma^{\pm}_j$ are generated by eigenvectors of  $\fA^p(\cdot, k)$ which  belong to none of the domains of pencils $\fM^q (\cdot, k)$, $q=1, \dots, N$.

For the solution $\cU$ in Proposition \ref{t3.1}, the radiation conditions (\ref{Radia1}) takes the form 
\begin{eqnarray}\label{RadiaM}
\cU - c_1 e^-_1-\dots -c_\Upsilon e^-_\Upsilon \in W^l_\delta (G); 
\end{eqnarray}
however the elliptic problem yet manifests itself  in the coefficients $c_j=i(\cF, \cV^-_j)_G$ in (\ref{RadiaM}), where $\cV^-_j$ is a solution to the homogeneous problem (\ref{ExtMax}), (\ref{ExtBound}). 
We have to show that  the role of $\cV^-_j$ is in fact played here by a solution to the homogeneous Maxwell system. It can be done by using the following
\begin{prop}\label{tescat}
There exists a unique solution $\cE_j^+ =(u^1, 0, u^2, 0)$ of the homogeneous problem {\rm (\ref{ExtMax}),
 (\ref{ExtBound})} such that 
\begin{equation}\label{EFCSM}
\cE_j^+(\cdot, k)- e_j^+ (\cdot, k)- \sum _{q=1}^\Upsilon s_{jq}(k)e^-_q(\cdot, k) \in W^1_\delta (G) 
\end{equation}
for $j=1, \dots, \Upsilon$. The functions $\cW^+_j: =(u^1, u^2)$ consisting of the components of $\cE_j^+$
form a basis in the space of continuous spectrum eigenfunctions of the Maxwell problem {\rm (\ref{Max}), (\ref{MaxBoun})}.
\end{prop}
In order to see that the scattering matrix $\sigma$ in Theorem \ref{t2.3} is block diagonal, we now combine Proposition \ref{tescat}
and the next
\begin{prop}\label{gscat}
There exists a unique solution $\cG_j^+$ of the homogeneous problem {\rm (\ref{ExtMax}), (\ref{ExtBound})}, such that 
\begin{equation}\label{gradFCS1}
\cG_j^+-\gamma^+_j-\sum\limits_{p=1}^{\Upsilon+N}\upsilon_{j p}\gamma^-_{p}\in W^1_\delta (G)
\end{equation}
for $j=1, \dots, \Upsilon+N$. The functions $\cE_1^+, \dots, \cE_\Upsilon^+, \cG_1^+, 
\dots, \cG_{\Upsilon+N}^+$ form a basis in the space of  continuous spectrum eigenfunctions  of problem {\rm (\ref{ExtMax}), (\ref{ExtBound})}.
The equality $\sigma  ={\rm diag}(s, \upsilon)$ holds, where $\sigma =\sigma (k)$ is the scattering matrix in Theorem \ref{t2.3}, 
$$
s=s(k)= (s_{jq}(k))_{j,\, q=1}^\Upsilon, \qquad \upsilon =\upsilon (k) =(\upsilon_{jq}(k))_{j,\, q=1}^{\Upsilon +N}.
$$
Since the matrix $\sigma$ is unitary, every block $s$ and $\upsilon$ is unitary as well.
\end{prop}

\end{document}